\documentclass{article}            
\usepackage{eufrak}
\begin{document}
\title{Dynamics of test bodies in scalar-tensor theory\\ and equivalence principle}

\author{Yuri N. Obukhov\\
\footnotesize \it Theoretical Physics Laboratory, Nuclear Safety Institute,\\
\footnotesize \it Russian Academy of Sciences, B.Tulskaya 52, 115191 Moscow, Russia\\
\footnotesize \it E-mail: obukhov@ibrae.ac.ru\\
\\
\\
Dirk Puetzfeld \\
\footnotesize \it ZARM, University of Bremen,\\
\footnotesize \it Am Fallturm, 28359 Bremen, Germany\\
\footnotesize \it E-mail: dirk.puetzfeld@zarm.uni-bremen.de, URL: http://puetzfeld.org}

\maketitle

\begin{abstract}
How do test bodies move in scalar-tensor theories of gravitation? We provide an answer to this question on the basis of a unified multipolar scheme. In particular, we give the explicit equations of motion for pointlike, as well as spinning test bodies, thus extending the well-known general relativistic results of Mathisson, Papapetrou, and Dixon to scalar-tensor theories of gravity. We demonstrate the validity of the equivalence principle for test bodies. 
\end{abstract}

{\footnotesize{\bf Keywords:} Scalar-tensor theories; Equations of motion; Equivalence principle.}

\section{Introduction}

Scalar-tensor theories are considered to be close and viable generalizations of Einstein's general relativity theory. Since their introduction in \cite{Jordan:1955,Jordan:1959,Thiry:1951} they attracted a lot of attention after the works of Brans and Dicke \cite{Brans:Dicke:1961,Brans:1962:1,Brans:1962:2,Dicke:1962:1,Dicke:1964} with the scalar field interpreted as a variable gravitational coupling -- for an overview of the history and results of scalar-tensor theories see \cite{Fujii:Maeda:2003,Brans:2005,Goenner:2012,Sotiriou:2014}.

However, little attention was paid to the study of motion of extended test bodies in such theories. Following \cite{Brans:1962:2,Bergmann:1968,Wagoner:1970}, the issue was thoroughly analyzed \cite{Damour:etal:1992} in the framework of the post-Newtonian formalism. Here we present the test body dynamics for a very large class of scalar-tensor theories. 

\section{Scalar-tensor theory}

We study a class of scalar-tensor theories (along the lines of \cite{Damour:etal:1992}) with the action $I = \int d^4x\,{\stackrel J{\mathfrak L}}$ on the spacetime manifold with the metric ${\stackrel Jg}{}_{ij}$. The Lagrangian density ${\stackrel {J}{\mathfrak L}} = {\stackrel J {\sqrt{-g}}}{\stackrel JL}$ has the following form
\begin{eqnarray}
{\stackrel JL} = {\frac 1{2\kappa}}\biggl(- F^2R({\stackrel Jg}) + {\stackrel Jg}{}^{ij}{\stackrel J\gamma}{}_{AB}\partial_i\varphi^A\partial_j\varphi^B - 2{\stackrel JU}\biggr) 
+ L_{\rm m}(\psi,\partial\psi,{\stackrel Jg}{}_{ij}).\label{LJ}
\end{eqnarray}
This extends the Brans-Dicke theory \cite{Will:1993,Will:2014} to the case with a multiplet of scalar fields $\varphi^A$ (capital indices $A,B,C = 1,\dots, N$ label the components of the multiplet). Here $\kappa = 8\pi G/c^4$ is Einstein's gravitational constant and 
\begin{equation}\label{AU}
F = F(\varphi^A),\qquad {\stackrel JU} = {\stackrel JU}(\varphi^A),\qquad {\stackrel J\gamma}_{AB} = {\stackrel J\gamma}{}_{AB}(\varphi^A).
\end{equation}
The Lagrangian $L_{\rm m}(\psi,\partial\psi,{\stackrel Jg}{}_{ij})$ depends on the matter fields $\psi$. 

The metric ${\stackrel Jg}{}_{ij}$ determines angles and intervals in the {\it Jordan reference frame}. The Riemannian curvature scalar $R({\stackrel Jg})$ is constructed from the Jordan metric. With the help of the conformal transformation
\begin{equation}
{\stackrel Jg}{}_{ij}\longrightarrow g_{ij} = F^2{\stackrel Jg}{}_{ij}\label{ggt} 
\end{equation}
we obtain the metric in the {\it Einstein reference frame}. 

In the Einstein reference frame the Lagrangian density in the scalar-tensor theory reads ${\mathfrak L} = \sqrt{-g}L$ with
\begin{eqnarray}
L = {\frac 1{2\kappa}}\left(- R + g^{ij}\gamma_{AB}\partial_i\varphi^A\partial_j\varphi^B - 2U\right) 
+ {\frac 1{F^4}}L_{\rm mat}(\psi,\partial\psi,F^{-2}g_{ij}).\label{LE}
\end{eqnarray}
The scalar curvature $R(g)$ is constructed from the Einstein metric $g_{ij}$, and
\begin{equation}
\gamma_{AB} = {\frac 1{F^2}}({\stackrel J\gamma}{}_{AB} + 6F_{,A}F_{,B}),\qquad 
U = {\frac 1{F^4}}{\stackrel JU}.\label{gamUt}
\end{equation}
The metrical energy-momentum tensor of matter is defined by $\sqrt{-g}t_{ij} := \\ 2 \partial (\sqrt{-g}{L}_{\rm mat}) / \partial g^{ij}$. The Noether theorem yields the conservation law
\begin{equation}\label{cons} 
\nabla_jt^{kj} = {\frac 1F}\left(4t^{kj} - g^{kj}g_{mn}t^{mn}\right)\partial_jF = - V_{ij}{}^kt^{ij}.
\end{equation}
For details see \cite{Obukhov:Puetzfeld:2014:2}. Here $V_{ij}{}^k = A_j\delta_i^k - {\frac 14}g_{ij}A^k$, and $A_i := \partial_i\log F^{-4}$.

\section{Equations of motion}

We derive the equations of motion in the Mathisson-Papapetrou-Dixon \cite{Mathisson:1937,Papapetrou:1951:3,Dixon:1964,Dixon:1974} approach by integrating the conservation law (\ref{cons}) using the geodesic expansion technique of Synge \cite{Synge:1960}. With the world function $\sigma$ and the parallel propagator by $g^{y}{}_{x}$, we introduce integrated moments to an arbitrary order $n=0,1,2,\dots$ by:
\begin{eqnarray}
p^{y_1 \dots y_n y_0}&:=& (-1)^n  \int\limits_{\Sigma(s)}\sigma^{y_1} \cdots \sigma^{y_n} g^{y_0}{}_{x_0}\sqrt{-g}t^{x_0 x_1} d \Sigma_{x_1}, \label{p_moments_def} \\
k^{y_2 \dots y_{n+1} y_0 y_1}&:=& (-1)^{n}  \int\limits_{\Sigma(s)} \sigma^{y_1} \cdots \sigma^{y_n} g^{y_0}{}_{x_0}g^{y_1}{}_{x_1}\sqrt{-g}t^{x_0 x_1} w^{x_2} d \Sigma_{x_2}.\label{k_moments_def}
\end{eqnarray}
We use a condensed notation so that $y_{n}$ denotes indices at the point $y$ which we associate with the world-line $y(s)$ of a test body, parametrized by the proper time $s$. The integrals are performed over spatial hypersurfaces $\Sigma(s)$.

\section{Pole-dipole equations of motion}

In the pole-dipole approximation, an extended body is characterized by the multipole moments $p^a, p^{ab}, k^{ab}, k^{abc}$. Using the general multipolar scheme \cite{Puetzfeld:Obukhov:2014} we derive the equations of motion for these moments:
\begin{eqnarray}
0 &=& k^{(a|c|b)} - v^{(a} p^{b)c}, \label{eom_2_n_2_ST} \\
\frac{D}{ds} p^{ab} &=&  k^{ba} - v^a p^b  - V_{dc}{}^b k^{acd},\label{eom_2_n_1_ST}\\ 
\frac{D}{ds} p^{a} &=& - V_{cb}{}^a k^{bc} - V_{dc}{}^a{}_{;b} k^{bcd} 
-\frac{1}{2} R^a{}_{cdb} \left(k^{bcd} + v^d p^{bc} \right).\label{eom_2_n_0_ST} 
\end{eqnarray}
Here $v^a:=dy^a/ds$ denotes the normalized four-velocity of a body. Since $k^{a[bc]} = 0$, we can solve (\ref{eom_2_n_2_ST}) to find explicitly
\begin{eqnarray}
k^{abc} = v^a p^{cb} + v^cp^{[ab]} + v^bp^{[ac]} + v^ap^{[bc]}.\label{kabcST}
\end{eqnarray}
Plugging this into (\ref{eom_2_n_1_ST}) and (\ref{eom_2_n_0_ST}), we obtain the generalized Mathisson-Papapetrou-Dixon system
\begin{eqnarray}
{\frac {D{\cal P}^a}{ds}} &=& {\frac 12}R^a{}_{bcd}v^b{\cal J}^{cd} - \xi\nabla^aF^{-4} - \xi^b\nabla_b\nabla^aF^{-4}, \label{DPtotST}\\
{\frac {D{\cal J}^{ab}}{ds}} &=& -\,2v^{[a}{\cal P}^{b]} - 2\xi^{[a}\nabla^{b]}F^{-4}.\label{DJtotST}
\end{eqnarray}
Here, following \cite{Puetzfeld:Obukhov:2013:3,Puetzfeld:Obukhov:2014,Obukhov:Puetzfeld:2014:2}, we introduce the generalized total energy--momentum 4-vector and the generalized total angular momentum by
\begin{eqnarray}
{\cal P}^a &:=& F^{-4}p^a + p^{ba}\nabla_bF^{-4},\label{PtotST}\\
{\cal J}^{ab} &:=& F^{-4}L^{ab}.\label{JtotST}
\end{eqnarray}
The orbital angular moment is defined by $L^{ab} := 2p^{[ab]}$, and we denoted
\begin{eqnarray}
\xi^a = -\,{\frac 14}g_{bc}k^{abc},\qquad \xi = -\,{\frac 14}g_{ab}k^{ab}.\label{xika}
\end{eqnarray} 

\section{Monopolar equations of motion}

At the monopolar order, the only nontrivial moments are $p^a$, and $k^{ab}$. The system (\ref{eom_2_n_2_ST})-(\ref{eom_2_n_0_ST}) then reduces to
\begin{eqnarray}
0 &=& k^{ba} - v^a p^b,\label{Mono1ST}\\
{\frac {Dp^a}{ds}} &=& -\,V_{cb}{}^ak^{bc}.\label{Mono2ST}
\end{eqnarray}
Making use of $k^{[ab]} = 0$, the first equation yields $v^{[a} p^{b]} = 0$, hence we have 
\begin{equation}
p^a = Mv^a\label{pMv}
\end{equation}
with the mass $M := v^ap_a$. Substituting (\ref{Mono1ST}) and (\ref{pMv}) into (\ref{Mono2ST}) we find $\xi = -\,{\frac {v^ap_a}4} = - \,{\frac M4}$ and 
\begin{equation}\label{MonoF}
{\frac {Dv^a}{ds}} = -\,(g^{ab} - v^av^b){\frac {\nabla_bF}{F}}. 
\end{equation}
Quite remarkably, the dynamics of an extended test body in the monopole approximation is independent of body's mass. For a trivial coupling function $F$, equation (\ref{MonoF}) reproduces the general relativistic result. Interestingly, the mass of a body is not constant: $M = F^3M_0$ with $M_0=$ const.

\section{Conclusions}

Our main result is the system (\ref{DPtotST})-(\ref{DJtotST}) that describes the dynamics of extended test bodies in scalar-tensor gravity. In the monopolar case, our analysis revealed a surprisingly simple equation of motion (\ref{MonoF}). In contrast to geodesic motion in General Relativity, freely falling massive test bodies in scalar-tensor gravity experience an additional force, determined by the new scalar degrees of freedom encoded in the function $F$. 

A remarkable feature of (\ref{MonoF}) is the prediction that all massive test bodies move in the same way, independently of their mass. We thus demonstrate the validity of the equivalence principle in scalar-tensor gravity. This is consistent with the previous observation \cite{Hui:2010,Hui:2009} that the total scalar charge of a body is equal to its mass when the scalar field self-interactions are neglected. The latter is in agreement with the test body assumption that underlies the Mathisson-Papapetrou-Dixon approach. When one goes beyond the test body approximation, however, the scalar charge is no longer equal to the mass and a further study is needed to fix their relation \cite{Gralla:2010,Gralla:2013}.

\paragraph{Acknowledgements}
D.P.\ was supported by the Deutsche Forschungsgemeinschaft (DFG) through the grant SFB 1128 (geo-Q).

\end{document}